\documentclass[aps,twocolumn,showpacs,preprintnumbers,amsmath,amssymb,floatfix]{revtex4}
\usepackage{graphicx}
\usepackage{epsfig}
\begin{document}
\def\be{\begin{equation}}
\def\ee{\end{equation}}
\def\bearr{\begin{eqnarray}}
\def\eearr{\end{eqnarray}}
\def\half{$\rm \frac{1}{2}$~}

\title{Theory of Confined High Tc Superconductivity in Monovalent Metals}

\author{G. Baskaran}
\affiliation
{The Institute of Mathematical Sciences, C.I.T. Campus, Chennai 600 113, India \&\\
Perimeter Institute for Theoretical Physics, Waterloo, ON N2L 2Y5, Canada}

\begin{abstract}
Monovalent non-transition metals are robust Fermi liquids. They defy superconductivity even at lowest temperatures (Li is a minor exception:Tc $\approx$ 0.4 mK). However, Thapa and Pandey \cite{ThapaPandey} have recently reported signals for ambient temperature granular superconductivity in Ag nanoparticle embedded in Au matrix. We develop a theory, where competing superconducing, CDW and SDW orders lose and get confined (go off-shell). They leave behind a robust Fermi liquid on-shell. \textit{A single half filled band crossing the Fermi level} provides a number of special k-space regions for \textit{singlet stabilizing umklapp pair scattering and superconductivity stabilizing repulsive pair scattering}. Carefully designed perturbations could deconfine a confined superconductivity. We suggest that electron transfer (doping) from Ag nanoparticles to Au matrix (with a higher electronegativity), quasi 2d structural reconstructions (e.g., 9R structure) at Ag-Au interfaces etc., bring out confined superconductivity. Beneath a calm Fermi sea, strong supercurrents may exist in several metals. 
\end{abstract}
 \maketitle
 
\section*{I. Introduction}

Superconductivity, a remarkable macroscopic quantum phenomenon, touches unexpected and different parts of physics and science in general. Its technological implications are also profound. Discovery of such a phenomenon in Hg in 1910, by Kamerlingh Onnes and decades of developments that followed have nurtured a strong desire and dream to realize superconductivity at ambient temperatures.

A milestone discovery by Bednorz and Muller \cite{BednorzMuller1986} of high Tc superconductivity in cuprates, in 1986, suggested that this dream could be realized. The resonating valence bond theory of high Tc superconductivity that began as a response to understand cuprate superconductors by Anderson \cite{PWA1987} and collaborators \cite{BZAandOthers} gave new hope and created a fertile ground for high Tc superconductivity, quantum spin liquids etc., in Mott insulators with and without doping.

We start with the well known fact that all non-transition monovalent metals resist superconductivity, even at the lowest temperatures (Li is an exception: Tc $\approx$ 0.4 mK). This is consistent with theory of phonon exchange and a large Coulomb pseudo potential $\mu^*$ \cite{PWAMorel}. Electronic mechanism of Kohn and Luttinger \cite{KohnLuttinger} and later works \cite{RaghuSteveDough,MaitiChubukov}, on systems with pure Coulomb interaction, give at best \textit{ultra low Tc's for equilibrium superconductivity,} in monovalent or jellium metals. Doped Mott insulator is a different story.

In this background a report of signals for ambient temperature granular superconductivity \cite{ThapaPandey} in Ag nanoparticles embedded in Au matrix has come as a surprise. 

We develop a theory, where we view Ag nano particles embedded Au matrix as  strongly perturbed monovalent metals. Strong perturbations liberate a \textit{confined superconductivity} in monovalent metals. Attempt is made to answer the following questions: i) what is the origin of hidden superconductivity and ii) how to find them ?

Briefly, because of half filled single band character of monovalent metals, Fermi surface offers good opportunity for spin singlet stabilizing quasi one dimensional like umklapp pair scattering processes. Even a modest repulsive U is of help. Zero momentum spin singlet repulsive Cooper pair scattering processes between regions surrounding special (L and X) points in the BZ take advantage of stabilized singlets and build certain latent superconducting orders: extended s-wave, spin singlet chiral and certain d-wave symmetries. We find the above, approximately, by constructing various k-space truncated Hubbard model.

How these orders remain latent is an interesting and difficult theoretical question. These orders compete among themselves and also with a strong CDW and SDW order a la' Overhauser. Every one loses and a robust Fermi liquid is left behind. The competition can be also viewed as a destructive quantum interference phenomena between 2 particle scattering processes in k-space.

We discuss the recently discovered \cite{ThapaPandey} signals of ambient temperature superconductivity signals, in the Ag nanoparticle embedded in Au matrix, in the light of our proposal. Strong perturbations such as electron transfer from Ag nanoparticles to Au matrix (arising from a difference in electronegativity between Ag and Au), quasi 2d structural reconstructions (e.g., 9R structure) at the Ag-Au interfaces etc., help stabilize a granular ambient temperature superconductivity, by discouraging competing orders. 

Au and Ag have a strong spin-orbit coupling. In our view spin orbit coupling present in Au and Ag do not play a decisive role in stabilizing ambient Tc superconductivity in the bulk. However, on the surface and interfaces one expects exciting topological features in superconductivity, arising from spin orbit coupling.

Our paper is organized as follows. After a brief summary of recent experimental observations we introduce a minimal model, namely a single band repulsive Hubbard model, at half filling, in fcc lattice with a small to moderate U. Then we discuss singlet stabilizing umklapp scattering. This is followed by a k-space truncated approach to find different hidden orders. It is followed by a discussion of how a latent high Tc superconductivity is brought out in the recent Ag-Au experiment, via certain strong perturbations. We end with a brief discussion that many metals may be harboring latent high Tc superconductivity and suggest some ways to study them and stabilise them.

\section*{I.a. Earlier Studies}

In the past there has been suggestions for various orders, particularly in monovalent alkali metals, such as Na and K. Overhauser has suggested a strong incommensurate CDW/SDW order arising from electron-electron repulsion and approximate Fermi surface nesting \cite{OverhauserBook}. 

Cohen \cite{CohenOddK} has invoked strong interaction between Landau quasi particles off the energy shell and suggested presence of odd k pairing in potassium, to explain Mayer-El Naby \cite{MayerNaby} high energy resonances at 0.8 eV in optical conductivity. The odd k character of superconducting order parameter $\Delta({\bf k})$ about the Fermi surface makes order parameter vanish all over the Fermi surface. Cohen called it as \textit{pairing of the second kind}. To the authors knowledge, this is the only work that discusses possibility of high energy off shell superconducting correlations in alkali metals.

On the experimental side, several anomalies exist in alkali metals, as elaborated and studied by Overhauser \cite{OverhauserBook} over years. There are also notable room temperature anomalies in Au, such as signals of ferromagnetism in Au nanoparticles \cite{AuFM} and anomalous diamagnetism in Au nanorods \cite{AuDiamagnetism} etc. 

\section*{I.b. Summary of ambient Tc Superconductivity in Ag-Au system.} 

In the work by Thapa and Pandey \cite{ThapaPandey}, nanoparticles of size $\sim$ 10 nm get embedded into a Au matrix. A sharp resistivity drop to values, below the scales of known best metals, at ambient temperatures is seen. There is also Meissner like diamagnetism that accompanies the resistivity transition. It is not a perfect diamagnetism - this has been attributed to granular superconductivity with a small Meissner fraction. It is also interesting that Tc is sensitive to external magnetic fields. In some samples, the above Meissner like signals are seen above room temperatures, indicating a great potential for increase of Tc.
	
It will be exciting to confirm these results, as they seem to open a new avenue in the search for room temperature superconductors. It will be beneficial to compare with various earlier and recent experimental claims of ambient temperature superconductivity in graphitic \cite{Esquinazi} and other \textit{elusive high Tc superconductors} \cite{Kopelevich}.

\section*{I.c. Known Examples of Fermi Liquids containing \\Latent Off Shell Orders}

In perturbative renormalization group approach \cite{ShankarRG} repulsive Hubbard U, in a single band Hubbard model in 3d, gets renormalized to zero and is irrelevant, when one starts with U much smaller than band width. This is the RG theory justification for validity of Fermi liquid theory for monovalent metals at low energy/temperature scales.

Modern entanglement renormalization \cite{GVidalMERA} perspective of quantum RG gives deep insights into nature of quantum fluctuations contained in the grounds state. We view it as a way or organizing and studying manybody quantum interference phenomena at different scales. Many interesting off shell phases and phenomena disappear, via k-space quantum interference, as one reaches low energy scales

We recall few examples, where competing order and new physics reveal themselves in a Fermi liquid as we move up in energy scale, or as we go off shell. Liquid He$^3$ is a good example. It is a Fermi liquid at very low energy scales. However, it bears interesting off-shell competing phenomena such as dynamical solid like short range correlations, exchange interaction and so on as seen in experiments. Well known heavy Fermions and Kondo systems are Fermi liquids below a very small, few meV scales. However, above this scale they become incoherent metals and non-Fermi liquids with rich physics of quantum entangled local moments and fluctuating valencies. Such crossover phenomena are generically difficult to handle theoretically.

\section*{II. Low Energy Electronic Model for Ag and Au}

Fermi surfaces of monovalent alkali metals are nearly spherical and are least affected by the underlying periodic pseudo potential. However, Cu, Ag and Au develop small necks at the BZ boundary, centered around L points. Though a small change, it modifies Fermi surface topology, from disjoint closed spherical like surfaces into connected surfaces in the extended BZ (reciprocal lattice is a bcc lattice).

What is important for us is that a single half filled band crosses the Fermi level in the first BZ. Further, a maximally localized Wannier function approach suggests a tight binding model for Cu and Au \cite{WFCuAgAu}. It has an s-like Wannier function for a single half-filled band that crosses the Fermi level and d like Wannier orbitals for the five filled d bands below the Fermi level. Metallic screening leaves behind a finite onsite Coulomb repulsion. 

Our theory for Cu, Au and Ag start with a simplified single orbital half filled band Hubbard model on an fcc lattice: 
\be
H = -t \sum_{\langle ij \rangle \sigma} (c^\dagger_{i\sigma}c^{}_{j\sigma} + H.c.) + U \sum_i n_{i\uparrow} n_{i\downarrow}
\ee

There is no particle-hole symmetry and one electron density of states is strongly asymmetric at half filling. Band width of Ag and Au are about 8 eV. Atoms in an fcc lattice have 12 nearest neighbors. Consequently, hopping matrix elements are small, t $\sim$ 0.5 eV. Hubbard U is in the range of 1 to 3 eV. It is small compared to band width.

\section*{II.a ~ Hubbard Model at Half Filling \& \\Study of Latent Singlet Superconducting Orders}

Single band Hubbard model in fcc lattice at half filling can be analysed using diagrammatic techniques or various RG schemes. There is a clear indication that low energy physics is governed by an unstable Fermi liquid fixed point, consistent with phenomenology. 

The Hubbard U term in momentum space is:
\bearr
U \sum_i n_{i\uparrow} n_{i\downarrow} &=& \frac{U}{N} \left(\sum_{k'} c^\dagger_{k'\uparrow} c^\dagger_{-k'\downarrow} \right) \left(\sum_k c^{}_{-k\downarrow}c^{}_{k\uparrow}\right) \nonumber \\
&+& {\rm umklapp~scattering~terms} \nonumber \\ 
&+&{\rm terms~with~CM~momentum \neq 0}
\eearr

First term is a positive definite. In a mean field picture for the ground state, the system would like to minimise this term. A simple meanfield solution, an absolute minima, is $ \Delta_s \equiv \sum_{k} \langle c^\dagger_{k\uparrow} c^\dagger_{-k\downarrow} \rangle$ = 0. This implies absence of simple s-wave pair condensation at a mean field level. 

Our thesis is that \textit{even though $\Delta_0$, an equal time, anomalous ground state average vanishes, large temporal fluctuations (in a range of frequency or energy scales) in the pair potential is present in various parts of k-space.}. In other words, a naive vanishing of $\Delta_0$ at mean field level could hide remarkable collective quantum interference phenomena in k-space,  

Proving this, calls for a careful study of frequency and momentum dependence of off diagonal two electron correlation functions for the present Hubbard model, which is technically challenging. 

\section*{II.b. Umklapp Pair Scattering \& Singlet Stabilization} 

To build a latent singlet superconductivity correlations, we need to produce strong correlations in appropriae frequency scales. In a singlet band Mott insulator (U large compared to band width) singlet formation is via antiferromagnetic superexchange. \textit{In the weak coupliing limit singlet pairs are formed via umklapp scattering}. That is, \textit{interaction induced dynamical singlets are formed  on top of kinematic singlets that exist in a Fermi sea, dictated by Pauli principle.}

{\bf Singlet Stabilization in One Dimension.} We illustrate this using known results for \textit{repulsive Hubbard model in 1d at half filling}, using Bosonization approach. Interestingly, \textit{spin singlet stabilization and Mott Hubbard gap formation are tied together}.

A pair of electrons with CM momentum -$\pi$ close to right Fermi point jumps to the left one and vice versa; CM momentum changes from $\pi$ to - -$\pi$ and vice versa. Total change in momentum in a jump is 2$\pi$, a reciprocal vector: this umklapp process, a violation of CM momentum by a reciprocal lattice vector is allowed in a lattice.
 
Mott Hubbard charge gap opens, via 2 electron umklapp pair scattering in spin singlet channel as follows. umklapp two particle scattering terms in the Hubbard interaction term takes a suggestive form \cite{Solyom}: 
\be
\sim U\int \cos\left( {\sqrt{8\pi}}~\phi_c(x,t)\right)~dx dt
\ee
Here $\phi_c$ is the charge phase field of the electron.  The cosine term locks the phase of spin zero electron pairs. This locking defines umklapp pair condensation or formation of charge neutral singlet pairs in a half filled band. We also recall that solitons of the charge phase field are the (spin zero) charge +e holon and charge -e doublon.

We should also point out that we have ignored forward scattering terms that are repulsive. At a mean field level this repulsive scattering exactly cancels the backward umklapp pair scattering; singlet stabilization does not take place. However, in an RG approach such a cancellation does not take place, because repulsive forward scattering gets renormalized downwards. This is an important feature of one dimensionality.

Above umklapp pair scattering leaves spin phase field $\phi_s$ free and gapless. Solitonic excitations of this free field (with compactification radius \half) produces spin-\half spinon excitations.

Thus umklapp scattering adds dynamic (interaction induced) singlets to the ground state. in 1d, it converts a charge Fermi sea into a neutral spinon Fermi sea and at half filling produces a charge gap..

{\bf Singlet Stabilization in Three Dimensions.} umklapp scattering have some intrinsic one dimensional character, even in the three dimensional case of interest to us, as we will see.

To construct low energy regions for umklapp pair scattering one defines umklapp surfaces.
These are pairs of planar surfaces in the BZ. They bound center of the BZ symmetrically.  Shortest vector between two planes of a pair is half a reciprocal lattice vector. Consequently, in momentum conserving two particle scattering, a jump from region near one plane to region near the parallel plane across is allowed. Violation of momentum conservation is by a reciprocal lattice vector, which is allowed in a lattice.

BZ of fcc lattice has 4 umklapp surface pairs parallel to hexagonal faces, and 3 pairs of surfaces parallel to square surfaces. It is easy to show that all 14 umklapp surfaces cut the Fermi surface of half filled band, on 14 distinct closed loops. These umklapp loops are \textit{hot lines}. analogue of hot spots in 2d. umklapp loops on the Fermi surface could cross each other. These crossing points are also special hot spots.

Consider a pair of parallel umklapp surface and low energy electron pairs states with center of mass momentum $\frac{G}{2}$, where G is the corresponding reciprocal lattice basis vector. We define a small energy shell of thickness $\epsilon_l$ at the chemical potential. Corresponding region, an umklapp shell $\Omega_G$is a torus in k-space that completely hides a given umklapp loop. We define an s-like combination of spin singlet pair operators carrying a CM momentum $\frac{G}{2}$ and $-\frac{G}{2}$ within two parallel umklapp shells:
\be
{\hat \Delta_G}^\dagger \equiv \sum_{k \in \Omega_G} c^\dagger_{k+\frac{G}{4}\uparrow}c^\dagger_{-k+\frac{G}{4}\downarrow}~{\rm and }~{\hat \Delta_{-G}}^\dagger \equiv \sum_{k \in \Omega_{-G}} c^\dagger_{k-\frac{G}{4}\uparrow}c^\dagger_{-k-\frac{G}{4}\downarrow}
\ee

Now we isolate two parallel umklapp loops and corresponding torii and keep repulsive scattering of a pair with CM momentum $\frac{G}{2}$, which takes a pair from one torus to the other. We drop rest of the pair scattering terms in the Hubbard model. This reduced Hubbard model 
\be
H_G = \sum_{k\sigma} \epsilon_k c^\dagger_{k\sigma}c^{}_{k\sigma} + \frac{U}{N}({\hat \Delta_G^\dagger} {\hat \Delta_{-G}^{}} + H.c )
\ee
supports Cooper pair condensation with CM momentum, $\frac{G}{2}$, half a reciprocal lattice vector. Here $\epsilon_k$ is the energy dispersion in an fcc lattice. Mean field values change sign between two parallel umklapp surfaces $\langle {\hat \Delta_{G}^{}}\rangle = - \langle {\hat \Delta_{-G}^{}}\rangle$, because of repulsive character of the scattering.

Without going to details, we would like to point out that umklapp scattering in 3d resembles a quasi 1d problem. Consequently the forward scattering (repulsive scattering within a umklapp torus) gets renormalized downwards. So singlet pair stabilization takes place from every pair of parallel umklapp surfaces.

There are a total of 7 umklapp surface pairs, each one contributes to singlet stabiization distributed on the Fermi surface. There are also special hot spots in 3 dimensions, where different umklapp loops meet. They also give rise to cross umklapp scattering between non parallel umklapp surfaces. 

Consequently umklapp scattering produces enough fluctuating dynamical singlets over an energy scale above the chemical potential. It is also important to note that umklapp pair scattering builds a dynamic Mott Hubbard correlation in a half filled band, even though we have a robust Fermi liquid at low energies.

Umklapp pair scattering is known to result in rich physics in 2d and stabilize a variety of phases \cite{umklappFukuyama} and also give a meaning to the preformed neutral singlet pairs of RVB state, in the pseudogap phase of the cuprates \cite{umklappRice}.

\begin{figure}
\includegraphics[width=0.3\textwidth]{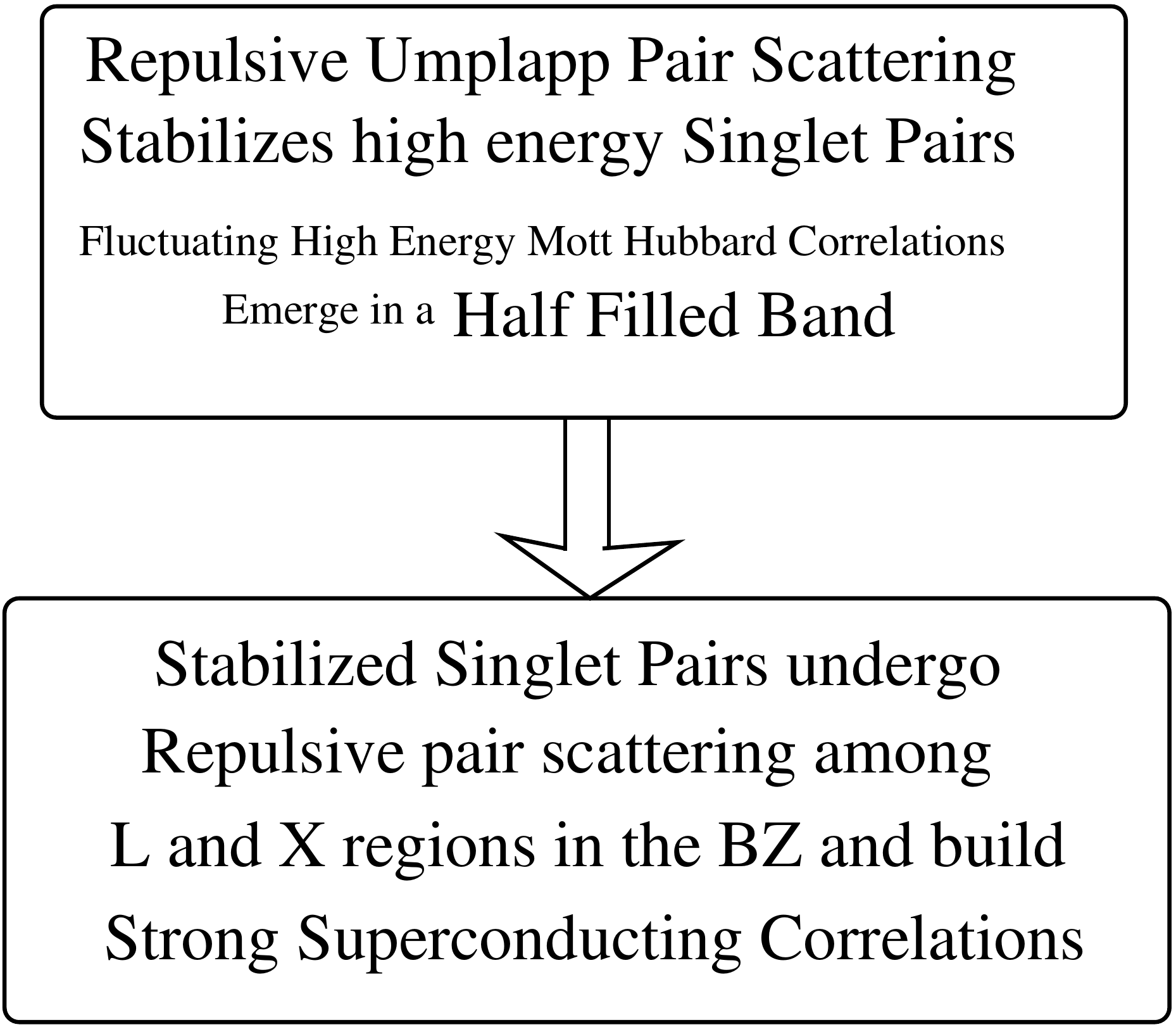}
\caption{Schematic illustration of growth of fluctuating singlet pairs by repulsive umklapp pair scattering and how they are being used by repulsive pair scattering among special regions near the Fermi surface, to build confined superconductivity.} \label{Figure 1}
\end{figure}

\section*{III. k-space Truncated Hubbard Model Approach}.
Having found processes that stabilize singlets on the Fermi surface we look for coherent scattering processes over the Fermi surface that could stabilize off shell or confined superconductivity (Figure 1).

Guided by physics, we simplify the Hubbard repulsive interactions in k-space, by systematically dropping all repulsive terms, except for a subset, and find various interesting mean field superconductivity solutions. \textit{Truncated Hubbard models in k-space, truncated in different ways reveal interesting hidden possibilities.}

Our approach is similar to keeping only the (repulsive) Ising term, and dropping the rest, in a study of 1d spin-\half Heisenberg antiferromagnetic chain. While this approximation misses presence of a spin liquid ground state, it definitely brings to the fore fluctuating antiferromagnetic order that is present off-shell, over a wide range of energy scales.

Important parts of the Fermi surface, from two particle scattering point of view in Ag and Au at half filling are regions around L points on the hexagonal face and X pockets on the square face on the boundary of BZ. We expect interesting physics from nearly resonant low energy (nearly elastic) scattering of electron pairs within these regions. In these regions nature of phase space (e.g., neck around L points, local 2d nature etc.) and density of states are different and are favorable for new emergent orders.

We choose a small energy shall of thickness $\epsilon_L$ and $\epsilon_X$, around four independent L and three independent X points. Shells are chosen to be symmetrical about the Fermi level. These are named $\Omega_l$ and $\Omega_x$ shells, with $l = l_1,l_2,l_3,l_4$ and $x = x_1,x_2,x_3$. 

It is easy to see that a $\Omega_l$ shell, a pocket in k-space, has the shape of a distorted torus that encloses a small cylindrical Fermi surface patch with open edges, in the neck region. On the other hand, $\Omega_x$ shell has two disconnected ellipsoidal pockets. They enclose two nearly circular and nearly parallel Fermi surface patches, which bound a given X point symmetrically (in an extended BZ).

Define an s-like combination of pair operators, within an energy shell, that carry a zero CM momentum and zero total spin:
\be
{\hat \Delta_l}^\dagger \equiv \sum_{k \in \Omega_l} c^\dagger_{k\uparrow}c^\dagger_{-k\downarrow}~~{\rm and}~~
{\hat \Delta_x}^\dagger \equiv \sum_{k \in \Omega_x} c^\dagger_{k\uparrow}c^\dagger_{-k\downarrow}
\ee

After a careful examination, we present three different k-space truncated Hubbard Hamiltonians containing only repulsive scattering terms. They are solvable at mean field level and exhibit, extended s-wave, PT violating chiral and d-wave order. In what follows we discuss only the first two and not the d-wave order.

In our truncated Hubbard Hamiltonian approach, k-space outside our chosen shells get decoupled. Hence we need to focus only on our $\Omega$ shells.

{\bf Extended s-wave Superconductivity.} Here we keep only inter shell scattering between 4 distinct L and 3 distinct X pockets. 

Notice that we have conveniently dropped repulsive scattering within a given $\Omega_l$ shell or $\Omega_x$ shell. As we discussed before, such terms destroy our mean field solution. However, arguments presented above, of proliferation of singlets all over the Fermi surface, via umklapp processes essentially renormalize intrashell scattering downwards, giving meaning to the mean field solution.

Order parameters on L shells and X shells have a relative sign change. We call them extended s-waves. In a more refined approach we will get nodal lines and sign changing order parameter on either side. It reminds one of Coulomb repulsion driven change of order parameter sign in two band model, found by Kondo \cite{Kondo2Band} and others.

{\bf Chiral Singlet Superconductivity.} We keep the three inter shell repulsive scattering among the 3 independent X shells and drop all the rest. In the minimum energy meanfield solution we get relative phase differences of either $e^{i\frac{\pi}{3}}$ or its complex conjugate $e^{-i\frac{\pi}{3}}$, in a cyclic order. These two degenerate solutions are parity and time reversal violating chiral spin singlet superconducting states.
Certain three patch interacting electron situations in two dimensions were found to have chiral spin singlet superconductivity \cite{chiral}.

{\bf Higher Angular Momentum Pairing.} We also get, via appropriate k-space truncation, a variety of confined spin singlet superconductivity with higher angular momentum pairings, with d-wave and g-wave symmetries. 

\section*{III.a. CDW and SDW Instability}

Possibility of CDW and SDW, arising from approximate Fermi surface nesting was extensively studied by Overhauser. He also discussed in great detail a variety of experimental results in favor of his proposal. In the present article we will not go into details, but to point to his monograph \cite{OverhauserBook}, \textit{Anomalous Effects in Simple Metals}, where dozens of detailed discussions are presented in support of incommensurate CDW and SDW orders.

\section*{III.b. Competing Orders, Fermi Liauid \&\\
Deconfinement of Confined Superconductivity}
It is clear experimentally that we have a robust Fermi liquid in all monovalent metals, at low energy scales. However, we saw in previous discussion that subsets of repulsive scattering, on their own can establish macroscopic quantum coherence via high Tc superconducting orders. Truncation we have implemented gets support from singlet stabilizing umklapp pair processes.

Quantum interference in k-space is at work: different coherent quantum processes interfere destructively. On top of this we have scattering of zero momentum Cooper pairs to finite CM momentum. Quantum interference takes place in the presence of Pauli blocking. Disentangling the collective quantum interference, hidden in a Fermi liquid, is a challenge problem.

A careful study of the destructive interference process and how we flow to a Fermi liquid fixed point requires a detailed RG study such as functional RG \cite{ShankarRG} or construction of tensor network states for the ground state \cite{GVidalMERA} and a proper interpretation. Then one will see the RG flow taking one close to various competing orders along the way. One can also study in detail their strength and their fluctuation energy scales, in principle. We hope to present this study in a future publication. Situation is reminiscent of taking care of interference via parquet diagram summation, when we have competing orders, in one dimensional and other contexts.
\begin{figure}
	\includegraphics[width=0.35\textwidth]{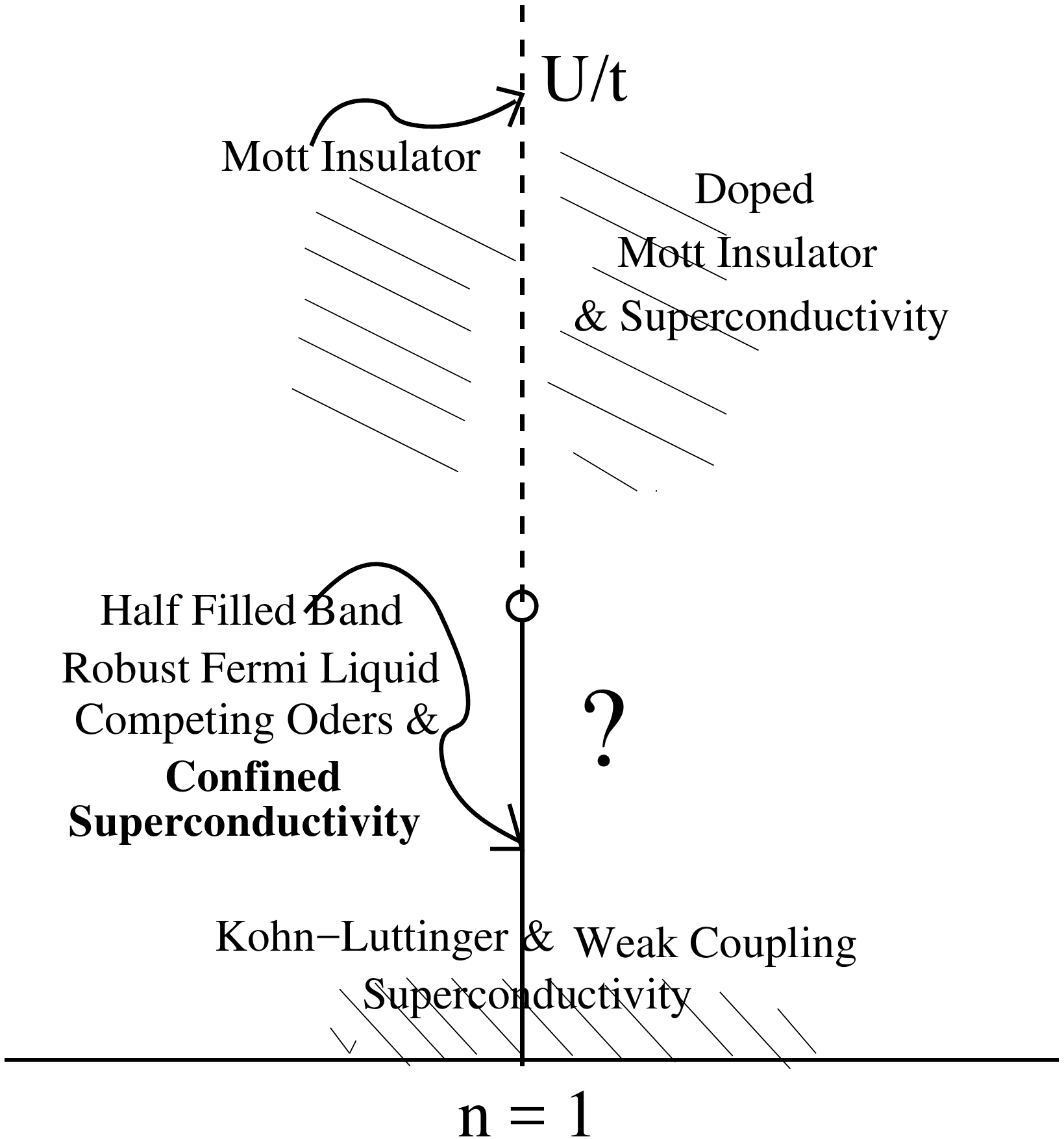}
	\caption{schematic phase diagram for repulsive Hubbard model on an fcc lattice, as a function of $\frac{U}{t}$ and site occupancy n. Question mark represents the region, where confined high Tc superconductivity could be deconfined; some additional help may be needed for deconfinement.} \label{Figure 1}
\end{figure}

Figure 2 presents a schematic phase diagram for repulsive Hubbard model on an fcc lattice. As we move away from half filling, in the region of confined superconductivity, there is an interesting question (marked by a question mark in figure 2). Do we have higc Tc superconductivity, via deconfinement of confined order, once competing superconducting and CDW/SDW orders lose out ? 

If we take existing phenomenology of monovalent metals, doping (for example alloying with divalent metal atoms) is not known to create high Tc superconductivity. It could mean that in addition to doping, other help such as structural reconstructions may be needed to deconfine the confined superconducting order. 

A careful theoretical study, including the often ignored umklapp singlet pair scatterin processes in weak coupling superconductvity study, is called for.

\section*{IV. Ag Nanocrystal in Au Matrix and Ambient Temperature Superconductivity}

In the Ag-Au composite one expects, a charge transfer from Ag to Au, close to the interface of nano crystal and bulk matrix. Electronegativity (ability to attract an electron) is higher for Au (2.54) than Ag (1.93). So we expect electron transfer from Ag to Au. Consequently, Ag will be hole doped and Au, electron doped. Metallic screening is likely to keep doping close to the interface.

Consequence of doping, or moving away from half filling, will discourage nesting instabilities. It will favor superconductivity.

Disorder in Ag-Ai system is present. We expect challenge for CDW and SDW arising from an inevitable atomic scale disorder at the Ag-Au interfaces. Similarly, chiral singlet will be strongly affected by disorder. Extended s-wave superconductivity will, however, be protected from disorder, via Anderson theorem as well as electron correlation effects. Thus doping, as a perturbation will encourage extended s-wave superconductivity.

Structural reconstructions are possible at the interface. Indeed reconstruction of fcc lattice to 9R, a quasi two dimensional structure has been seen in experiment \cite{Ag9R} for Ag and Cu. Further, theoretical studies of 9R structure for Li \cite{LiHoffmann} a monovalent metal, shows a quasi 2d Fermi surface. An emergent local 2d Fermi surface will also support superconductivity.

It is also known that Ag and Au (111) surfaces support 2d surface bands. Do they or a hybridized version them survive at the Ag nanocrystal Au interface, and support high Tc superconductivity ?
 
As scale of hopping matrix element is high about 0.5 eV, singlet pairing strength and Cooper pairing strength are high. In our estimate they are in the right ball park to reach ambient temperature superconductivity, seen in experiments.

As we discussed before, gapless extended s-wave, gapful chiral singlet instabilities compete. At the moment extended s-wave seems to be the winner. However, chiral singlet order, d and g-wave can not be excluded. Further xperiments and theories can be a good guide. 

\section*{IV.a. Scales of Energy Gap of\\ Confined Superconductivity \\ \& Possible Experimental Detection}
In our mean field solutions of truncated Hubbard model, by a proper choice of the energy shell we get very large energy scale for the superconducting gaps. As we are analysing only the bare truncated Hamiltonian, without any renormalization, we get embarassingly large values for Tc. Our present theory indicates possibility of ambient temperature superconductivity. More refined analysis is necessary for quantitative purposes.

It is also very exciting that the Mayer-El Naby \cite{MayerNaby}
high energy resonance in potassium gives a large energy gap (0.3 eV), interpreted as an odd-k pairing gap by Cohen \cite{CohenOddK} and CDW gap by Overhauser \cite{OverhauserBook}. It will be interesting to perform optical conductivity, high energy neutron scattering, EELS or other experiments and look for signals for \textit{high energy fluctuations or confined orders} in Ag and Au.

\section*{V. Discussion}
Inspired by the recent discovery of strong signals for ambient temperature superconductivity in Ag nanoparticle embedded in Au matrix, we have presented a theoretical analysis in support of their discovery. Our main thesis is that a robust Fermi liquid, in view of their half filled single band character, actually supports competing orders, including high Tc superconductivity. 

In pure elemental metals, high Tc superconductivity loses via quantum interference of (superconductivity, CDW and SDW stabilizing) scattering processes in k-space. Carefully chosen perturbations can make high Tc superconductivity win in monovalent metals. 

Our proposal also suggests that other composites involving two are more of monovalent metals could give rise to ambient Tc superconductivity. 

Another interesting approach suggested by our work is non equilibrium or driven high Tc superconductivity. A driven current or femto second pulses may capture off shell incipient pairing and create transient order \cite{Cavalleri}.

Our analysis may be relevant in general metals, with complex Fermi surfaces that support umklapp pair scattering

To confirm our proposal it will be very interesting to study isolated nanocrystals, quantum dots and rods of monovalent metals, using STM and other probes, after surface or bulk doping, or gate doping. Study of high frequency charge 2e noise will also be valuable. Large signals of diamagnetism, seen \cite{AuDiamagnetism} in Au nano rods etc., at ambient temperatures are exciting signals, from the point of view of our present proposal. 

Fermi liquids abound in metal physics. Our theory and truncated Hamiltonian approach and quantum interference in k-space suggests systematic investigations of other monovalent systems and nano metal composites. 
\section*{Acknowledgement}

I thank R. Ganesh, R. Shankar and Venky Venkatesan for discussion. I acknowledge an early (2003) discussion with (late) A. Overhauser at Purdue on his ideas on anomalies in simple alkali metals, and his insistence that we should not ignore them. A Distinguished Fellowship from Science and Engineering Research Board (SERB, India) is acknowledged. It is a pleasure to thank Perimeter Institute for Theoretical Physics (PI), Waterloo, Canada for a DVRC and hospitality. PI is supported by the Government of Canada through Industry Canada and by the Province of Ontario through the Ministry of Research and Innovation.

\end{document}